\documentclass[aps,prl,showpacs,twocolumn]{revtex4}
\usepackage{amsmath}
\usepackage[dvips]{graphicx}
\usepackage{latexsym}
\usepackage{amsfonts}
\usepackage{amssymb}

\usepackage{colordvi}
\usepackage{color}

\newcommand{\ket}[1]{\ensuremath{\left| #1 \right>}}

\begin{document}

\title{A high-fidelity noiseless amplifier for quantum light states}

\author{A. Zavatta$^{1,2}$, J. Fiur\'{a}\v{s}ek$^{3}$, and M. Bellini$^{1,2}$}

\affiliation{$^1$Istituto Nazionale di Ottica (INO-CNR), L.go E. Fermi 6, 50125 Florence, Italy\\
$^2$LENS and Department of Physics, University of Firenze, 50019 Sesto Fiorentino, Florence, Italy\\
$^3$Department of Optics, Palack\'{y} University, 17. listopadu 12, 77146 Olomouc, Czech Republic}

\date{\today}


\begin{abstract}
Noise is the price to pay when trying to clone or amplify arbitrary quantum states. The quantum
noise associated to linear phase-insensitive amplifiers can only be avoided by relaxing the
requirement of a deterministic operation. Here we present the experimental realization of a
probabilistic noiseless linear amplifier that is able to amplify coherent states at the highest
level of effective gain and final state fidelity ever reached. Based on a sequence of photon
addition and subtraction, and characterized by a significant amplification and low distortions,
this high-fidelity amplification scheme may become an essential tool for quantum communications and
metrology, by enhancing the discrimination between partially overlapping quantum states or by
recovering the information transmitted over lossy channels.
\end{abstract}

\maketitle

Quantum mechanical laws impose that arbitrary quantum states cannot be perfectly cloned or amplified
without introducing some unavoidable noise in the process \cite{Caves82,Scarani05}. This is a
consequence of the linearity and unitary evolution of quantum mechanics and guarantees against
unphysical situations such as the violation of the Heisenberg uncertainty principle or the superluminal
exchange of information \cite{Gisin98,Bruss00}. In any deterministic linear amplifier noise is
unavoidably added in the process and any input pure state results in a mixed output one.

This has profound implications also from a practical point of view in the frame of quantum information
processing and quantum metrology. For example, it greatly limits the possibility of restoring the
information carried by some fragile quantum state by amplifying it after it has been degraded in a lossy
channel. Or, it can forbid to distinguish among different parameter values if they are encoded in
partially overlapping quantum states.

As an illustrative example let us consider the case that some quantum information (or classical
parameter value) is encoded in the complex amplitude $\alpha$ of a coherent state $\ket{\alpha}$. If the
state amplitude is made too small (generally by losses) then the strong overlap between different states
can make it impossible to correctly distinguish among them. Simply amplifying the states would not solve
the problem because it would also amplify the quantum fluctuations of the coherent states, thus in fact
increasing their overlap and making the situation even worse (see Fig. 1).
\begin{figure}[h]
\includegraphics [width=0.8\linewidth]{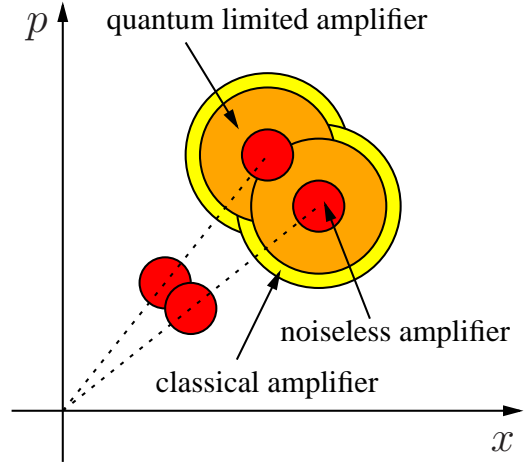}
\caption{Wigner function contours of input and amplified coherent states. The quantum-noise limited
amplifier with amplitude gain $g$ invariably adds noise being at least the equivalent of $2(g^2-1)$
vacuum-noise units. The best classical linear amplifier based on measure-and-prepare strategy adds even
more noise, namely at least $2g^2$ vacuum-noise units. By contrast, the probabilistic noiseless
amplifier preserves the noise of coherent states while amplifying their amplitude.}
\end{figure}

A solution to this problem would be provided by an ideal noiseless amplifier of coherent states of light
whose action can be mathematically described as
\begin{equation}
|\alpha\rangle \rightarrow |g\alpha\rangle, \label{amplification}
\end{equation}
where $g>1$ is the amplification gain. Referring to the above example, a sufficient noiseless
amplification of partially-overlapped coherent states would allow one to make them exactly
distinguishable.

The transformation (\ref{amplification}) is unphysical, but can be implemented probabilistically in an
approximate way. Ralph and Lund \cite{Ralph08} recently proposed a scheme based on the application of
multiple quantum-scissors blocks \cite{Pegg98,Babichev03} to non-deterministically amplify the
low-amplitude portions obtained by splitting a coherent state before their coherent recombination in an
interferometric setup. Although the complete scheme is almost impossible to realize with current
technologies, the functioning of its quantum-scissors core element has been recently demonstrated by two
experimental groups \cite{Xiang09,Ferreyrol10}.

Here we follow a completely different route, based on a combination of photon addition and subtraction,
and show that the performances of this approach are far superior, both in terms of higher effective
amplification, and of higher fidelity of the final states to the ideal target coherent state
$|g\alpha\rangle$.

Addition and subtraction of single photons are the result of the application of the creation and
annihilation operators $\hat{a}^\dagger$ and $\hat{a}$ to an arbitrary state of light. Depending on the
ordering of such operations, a transformation $\hat{a} \hat{a}^\dagger$ or $\hat{a}^\dagger \hat{a}$ can
be applied to the initial state. Sequences and coherent superpositions of such quantum operators have
been recently demonstrated experimentally \cite{Parigi07,Zavatta09}. Making a coherent linear
combination of these two operations with suitable weights one can obtain
\begin{equation}
\hat{G}=(g-2)\hat{a}^\dagger \hat{a} + \hat{a} \hat{a}^\dagger = (g-1)\hat{n}+1, \label{G}
\end{equation}
where $\hat{n}=\hat{a}^\dagger \hat{a}$ is the photon number operator. As shown in Ref.
\cite{Fiurasek09}, the operation (\ref{G}) is a good approximation of the ideal noiseless
amplification process (\ref{amplification}) for weak coherent states.

The performance of the approximate amplifier (\ref{G}) can be quantified by its effective gain and
fidelity. For an input coherent state $|\alpha\rangle$, the un-normalized output state of the amplifier
reads $\hat{G}|\alpha\rangle$. We define the effective amplification gain $g_{\mathrm{eff}}$ as the
ratio of the mean values of annihilation operator $\hat{a}$ for the output state $\hat{G}|\alpha\rangle$
and input state $|\alpha\rangle$. Since $\langle \alpha |\hat{a}|\alpha\rangle=\alpha$ we have
\begin{equation}
g_{\mathrm{eff}}= \frac{1}{\alpha} \frac{\langle \alpha|\hat{G} \hat{a}\hat{G}|\alpha\rangle}{\langle
\alpha|\hat{G}^2|\alpha\rangle}. \label{geffdefinition}
\end{equation}
On inserting the expression for the operator $\hat{G}$ into Eq. (\ref{geffdefinition}) we obtain after
some algebra
\begin{equation}
g_{\mathrm{eff}}=1+\frac{(g-1)\left[1+(g-1)|\alpha|^2\right]}{1+(g^2-1)|\alpha|^2+(g-1)^2|\alpha|^4}.
\label{geff}
\end{equation}
The fidelity of the amplifier is defined as normalized overlap of the output state
$\hat{G}|\alpha\rangle$ with the ideal target coherent state $|g\alpha\rangle$,
\begin{equation}
F=\frac{|\langle g\alpha|\hat{G}|\alpha\rangle|^2}{\langle \alpha|\hat{G}^2|\alpha\rangle}.
\label{Fdefinition}
\end{equation}
A straightforward calculation yields
\begin{equation}
F=\frac{\left[1+g(g-1)|\alpha|^2\right]^2
e^{-(g-1)^2|\alpha|^2}}{1+(g^2-1)|\alpha|^2+(g-1)^2|\alpha|^4}. \label{F}
\end{equation}

Of particular interest is the nominal gain $g=2$. In this case the formula for $\hat{G}$ simplifies, as
one term in the superposition (\ref{G}) vanishes and we obtain
\begin{equation}
\hat{G}_{g=2}=\hat{a}\hat{a}^\dagger. \label{Gtwo}
\end{equation}
Application of such transformation for noiseless amplification has been originally discussed in Ref.
\cite{Marek10}, and its action is evident if applied to a weak coherent state approximately described as
$\ket{\alpha}=\ket{0}+\alpha \ket{1}$: one gets $\hat a \hat a^{\dag} (\ket{0}+\alpha
\ket{1})\rightarrow \hat a (\ket{1}+\sqrt{2}\alpha \ket{2})\rightarrow \ket{0}+2\alpha \ket{1}$, i.e., a
doubling of the coherent state amplitude. The advantage of the transformation (\ref{Gtwo}) is that its
experimental implementation does not require interferometric stability unlike the general case of $g
\neq 2$.

The experiment is based on a unique and versatile setup for implementing creation and annihilation
operators that has been recently used to arbitrarily engineer quantum light states and test fundamental
quantum mechanical rules \cite{Parigi07,Kim08,Zavatta09,zavatta04:science}. The addition of a single
photon to an arbitrary light state is obtained by conditional stimulated parametric down-conversion in a
nonlinear crystal. The photon addition in the output signal mode is heralded by the detection (by an
on/off photodetector $D_a$) of a single photon in the idler down-conversion channel. On the other hand,
single-photon subtraction is implemented by conditionally attenuating a state by detecting (by an on/off
photodetector $D_s$) a single photon reflected from a high-transmissivity beam-splitter (BS). By placing
the parametric down-converter and the beam-splitter in series along the path of a traveling coherent
state, one can herald the application of the $\hat a \hat a^{\dag}$ operator by looking for coincident
detections from $D_a$ and $D_s$, as shown in Fig. 2. The low parametric gain and the low reflectivity of
BS (set to about 5 $\%$ for these measurements) guarantee that the experimental scheme is a very
faithful implementation of the ideal operator sequence.
\begin{figure}[h]
\centerline{\includegraphics[width=0.9\linewidth]{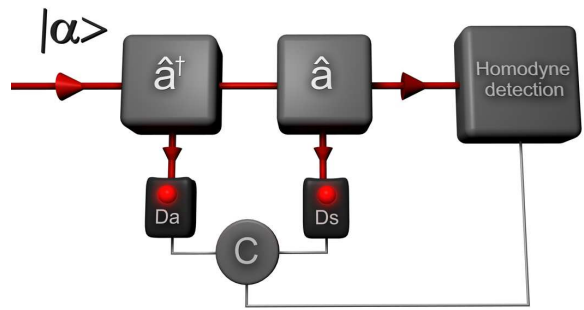}} \caption{Schematic experimental
setup. Two blocks for conditional single-photon addition and subtraction are placed in the path of a
coherent state. A coincident click (C) from the two on/off photodetectors heralds the successful
realization of the $\hat a \hat a^{\dag}$ operator sequence and the probabilistic noiseless
amplification of the input coherent state. High-frequency, time-domain, balanced homodyne detection is
then used for a full reconstruction of the involved quantum states.}
\end{figure}

The main light source is a mode-locked Ti:sapphire laser producing 1.5 ps pulses at 786 nm and with a
repetition rate of about 82 MHz. Most of the laser emission is frequency-doubled to become the pump for
the down-conversion process. An attenuated portion of the laser emission is used as the coherent field
$\ket{\alpha}$, which is injected along the signal mode of the down-converter (a type-I, $\beta-$barium
borate crystal) and eventually crosses a variable-reflectivity beam-splitter (BS, a half-wave plate and
polarizing beam-splitter combination) before being mixed in a 50-50 beam-splitter with another portion
of the original laser field serving as the local oscillator (LO) for balanced time-domain homodyne
detection. Differently from previous experiments that only involved single-click heralding or
phase-independent states, here a particular care has to be taken in order to perform phase-sensitive
homodyne measurements triggered at a relatively low rate (ranging from about 20 cps for low $|\alpha|$
values to about 70 cps for $|\alpha|=1$). An active stabilization of the relative phase between the
signal state and the LO has been implemented to this purpose by using the DC component of the homodyne
current as a control signal in a feedback loop.

Quadrature measurements for the amplified $\hat a \hat a^{\dag}\ket{\alpha}$ state are obtained by time
integration of the pulsed homodyne signal synchronous to a coincident $D_a-D_s$ click. The next pulses
(not coincident with any trigger event) are also analyzed in order to acquire homodyne data for the
corresponding un-amplified $\ket{\alpha}$ input state. Finally, quadrature measurements of the vacuum
state obtained by blocking the signal beam are also acquired for normalization. An absolute calibration
of the input coherent state amplitude $|\alpha|$ is obtained by comparing the rate of stimulated photon
addition events to spontaneous ones.

The experimental estimation of the effective gain $g_{\mathrm{eff}}$ is simply obtained by measuring the
ratio of the mean values of the amplitude quadratures for the output and the input states. This is done
by locking the relative phase between the coherent state $\ket{\alpha}$ and the local oscillator to an
interference maximum (or minimum). Interestingly enough, since these two quantities are measured with
the same homodyne detector, they suffer identical losses, therefore detection inefficiency factors out
in their ratio. The experimental effective gain is plotted in Fig. 3A as a function of $|\alpha|$,
together with that calculated for $g=2$. For low values of $|\alpha|$ the effective gain is very close
to the ideal value of 2, corresponding to an effective intensity amplification of $\approx4$.
\begin{figure}[h]
\includegraphics[width=0.9\linewidth]{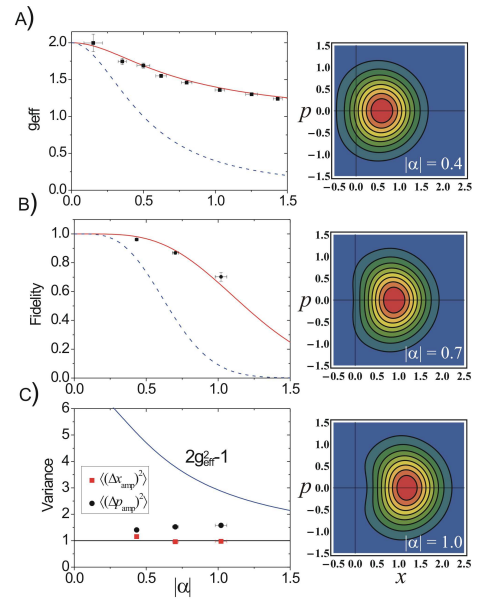}
\caption{Dependence of the (A) effective gain and (B) final state fidelity vs. input state amplitude
$|\alpha|$ for a nominal gain $g=2$. Red solid curves are calculated for the addition/subtraction
scheme; blue dashed curves are for the quantum-scissors method; square dots indicate experimental data.
(C) Measured variances (corrected for the detection efficiency $\eta=0.6$) of the amplitude and phase
quadratures of the amplified coherent state and the corresponding (blue solid) curve for the best
deterministic amplifier. The right panels show contour plots of the reconstructed Wigner functions for
three amplified coherent states of different amplitudes.}
\end{figure}

About $10^5$ quadrature measurements distributed in 11 values of the LO phase in the [0, $\pi$] interval
are then acquired to perform a quantum tomographic reconstruction of the states based on an iterative
max-likelihood algorithm \cite{lvovsky04,hradil06}. The experimental fidelity of the amplified state to
the target state is calculated by comparing the amplified state to a coherent state of double amplitude
$\ket{2\alpha}$ (obtained by halving the amplitude attenuation experienced by the portion of the laser
emission injected in the down-converter crystal), through their reconstructed density matrix elements.
Experimental fidelity values corresponding to three different amplitudes of the input coherent state are
plotted together with the calculated curves in Fig. 3B. We find a very good agreement with the expected
behavior and a high-fidelity ($F>90\%$) operation of our noiseless amplifier is preserved up to input
coherent state amplitudes $|\alpha|\lesssim 0.65$, corresponding to $g_{\mathrm{eff}}\approx 1.6$. The
very high fidelity of our noiseless amplifier is also evident in the little distortions experienced by
the Wigner functions of the amplified states, whose contour plots are also shown in the right panels of
Fig. 3.

The noise properties of the amplifier may be succinctly characterized by measuring the variances of the
amplitude and phase quadratures ($x_{\mathrm{amp}}$ and $p_{\mathrm{amp}}$) of the amplified coherent
state. The results are plotted in Fig. 3C, where one can see that the variances lie far below the value
of $2g_{\mathrm{eff}}^2-1$ shot-noise units corresponding to the best deterministic linear amplifier.
From the knowledge of the quadrature fluctuations and the effective amplification gain we can also
determine the equivalent input noise of the amplifier \cite{Roch93,Grosshans03,Ferreyrol10},
\[
N_{\mathrm{eq}}=\frac{\langle (\Delta x_{\mathrm{amp}})^2 \rangle }{g_{\mathrm{eff}}^2} -\langle (\Delta
x_{\mathrm{in}})^2 \rangle.
\]
A direct calculation reveals that the approximate noiseless amplification $\ket{\alpha} \rightarrow a
a^\dagger \ket{\alpha}$ exhibits negative $N_{\mathrm{eq}}$ for all $\alpha$. Experimentally, we find
that our amplifier indeed achieves $\mathrm{N}_{\mathrm{eq}}<-0.48$ for all considered coherent state
amplitudes $|\alpha|\leq 1.4$.

It is now quite instructive to compare the performances of the amplifier based on the combination of
photon addition and subtraction to those of other schemes of noiseless amplification. In the experiments
based on quantum scissors \cite{Xiang09,Ferreyrol10}, the state is truncated at Fock state $|1\rangle$,
whose weight is increased so as to emulate the amplification. An output state of such amplifier
corresponding to the input coherent state $|\alpha\rangle$ is thus given by
\begin{equation}
|\psi(\alpha)\rangle=\frac{1}{\sqrt{1+g^2|\alpha|^2}} (|0\rangle+g\alpha| 1\rangle).
\label{psidefinition}
\end{equation}
The effective gain and fidelity of amplifier based on quantum scissors can be defined similarly as
above, only the output state $\hat{G}|\alpha\rangle$ has to be replaced with $|\psi(\alpha)\rangle$,
\begin{equation}
g_{\mathrm{eff,QS}}=\frac{1}{\alpha}\langle \psi(\alpha)|\hat{a}|\psi(\alpha)\rangle, \qquad
F_{QS}=|\langle g\alpha|\psi(\alpha)\rangle|^2.
\end{equation}
On inserting the state (\ref{psidefinition}) into these formulas we obtain
\begin{equation}
g_{\mathrm{eff,QS}}=\frac{g}{1+g^2|\alpha|^2}, \label{geffQS}
\end{equation}
and
\begin{equation}
F_{QS}=\left(1+g^2|\alpha|^2\right)e^{-g^2|\alpha|^2}. \label{FQS}
\end{equation}

For $g=2$, we obtain from Eqs. (\ref{geff}), (\ref{F}), (\ref{geffQS}) and (\ref{FQS}) the following
expressions,
\begin{equation}
g_{\mathrm{eff}}=1+\frac{1+|\alpha|^2}{1+3|\alpha|^2+|\alpha|^4}, \qquad
g_{\mathrm{eff,QS}}=\frac{2}{1+4|\alpha|^2},
\end{equation}
and
\begin{equation}
F=\frac{\left(1+2|\alpha|^2\right)^2 e^{-|\alpha|^2}}{1+3|\alpha|^2+|\alpha|^4}, \qquad
F_{QS}=(1+4|\alpha|^2) e^{-4|\alpha|^2}.
\end{equation}
The effective gain and fidelity of the amplifier based on quantum scissors for a nominal gain $g=2$ are
also plotted in Figs. 3A and 3B. Both quantities decrease with increasing $|\alpha|$ but the amplifier
based on the combined photon addition and subtraction greatly outperforms the one based on quantum
scissors, also in terms of equivalent input noise. In fact, the scissors-based amplifier can even lead
to effective attenuation, $g_{\mathrm{eff}}<1$, because the very crude state truncation becomes the
dominant effect as soon as the condition $|g\alpha|^2 \ll1$ is not satisfied. The fidelity of the
amplification achieved with our approach is also much better than the one achievable by the scheme based
on thermal noise addition and single-photon subtraction proposed in \cite{Marek10}. This scheme is
unavoidably limited in its performance and can exhibit high fidelity only in the regime where the
effective gain drops very quickly with increasing $|\alpha|$. Although still able to conditionally
improve phase estimation, the thermal addition has the detrimental side-effect of significantly reducing
the purity of the amplified state, whereas our scheme is in principle able to preserve the unit purity
of the input coherent states.

Among other applications, noiseless amplification can enhance the performance of state-discrimination
and phase-estimation schemes. In particular, consider a protocol where a unitary transformation
$|\alpha\rangle \rightarrow |\alpha e^{i\theta}\rangle$ imprints information about phase shift $\theta$
onto the phase quadrature $p$ that is measured by a balanced homodyne detector. We have $\langle p
\rangle= 2|\alpha|\sin\theta$, and for small $\theta$ we may construct an estimator
$\theta_{\mathrm{est}} = \frac{p}{2|\alpha|}$ whose variance is inversely proportional to the total mean
number of photons in the probe coherent state, $V(\theta_{\mathrm{est}})=\frac{1}{4|\alpha|^2}$, which
is the well-known standard quantum limit \cite{Dowling98,Giovannetti04}. If the coherent state is
noiselessly amplified before detection, the variance of $\theta_{\mathrm{est}}$ is conditionally reduced
by a factor $R_V=g_{\mathrm{eff}}^{-2}\langle (\Delta p_{\mathrm{amp}})^2 \rangle/\langle (\Delta
p_{\mathrm{in}})^2\rangle$. For a perfect noiseless amplifier one gets $R_V=g_{\mathrm{eff}}^{-2}$.
Experimental values of $R_V=0.45, 0.64, 0.76$ for $|\alpha|=0.4, 0.7, 1.0$, respectively, indicate the
clear improvement in phase estimation achieved with the present scheme. The state-discrimination and
phase-estimation ability of our amplifier is further illustrated in Fig. 4, where the Wigner function of
an incoherent mixture of two coherent states with the same amplitude $|\alpha|=1.0$ and a $\pi/2$ phase
offset is shown before and after noiseless amplification by the photon addition and subtraction scheme.
\begin{figure}[h]
\includegraphics[width=.9\linewidth]{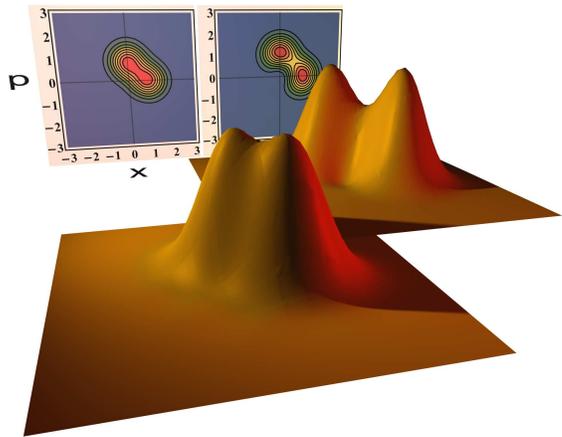}
\caption{Experimental Wigner functions for an incoherent mixture of $\ket{\alpha}$ and $\ket{i\alpha}$
before (front) and after (rear) amplification, with $|\alpha|=1.0$. The equal-weight incoherent mixtures
are simulated by summing the experimentally-reconstructed Wigner functions and those obtained by
imposing them a $\pi/2$ phase shift.}
\end{figure}
The effect of our high-fidelity noiseless amplifier is that of allowing a clear discrimination and a
much better phase estimation for the states that were almost totally overlapped before amplification.

We anticipate numerous applications of the demonstrated noiseless amplifier in quantum information
processing and quantum metrology. It can compensate for losses in quantum communication schemes and can
be used to distill and concentrate entanglement \cite{Ralph08,Fiurasek09}. Since it preserves quantum
coherence it could be used for breeding small cat-like states of the form $\ket{\alpha} \pm
\ket{-\alpha}$. As clearly shown above, it can improve the performance of phase-estimation schemes
\cite{Marek10} and enable high-fidelity probabilistic cloning and discrimination of coherent states.
Moreover, a fully tunable amplification gain can be achieved with an extended interferometric version of
the present setup that can also emulate Kerr nonlinearity \cite{Fiurasek09}. The present approach to
high-fidelity noiseless amplification, largely outperforming concurrent schemes, will certainly
represent an essential tool for the emerging quantum technologies.

A.Z. and M.B. acknowledge support by Ente Cassa di Risparmio di Firenze and Regione Toscana under
project CTOTUS. J.F. acknowledges support by MSMT under projects LC06007, MSM6198959213, and 7E08028, by
EU under the FET-Open project COMPAS (212008) and by GACR under project GA202/08/0224.

\bibliography{Fock_bib}

\end{document}